\documentclass[10pt,twocolumn,twoside]{IEEEtran}

\usepackage[dvips]{graphicx}
\usepackage{epsf}
\usepackage{epsfig}
\usepackage{psfrag}
\usepackage{amsmath, alltt, amssymb, xspace, color, cite, tabu}
\usepackage{subfigure}
\usepackage{tikz}
\usepackage{stfloats}

\usetikzlibrary{arrows,automata}
\usetikzlibrary{positioning}
\usetikzlibrary{shapes}

\def\Phib{{\mathbf \Phi}}

\newtheorem{assumption}{Assumption}
\newtheorem{lemma}{Lemma}

\newtheorem{theorem}{Theorem}

\begin{document}

\title{Fusion-Based Cooperative Support Identification for Compressive Networked Sensing}

\author{Ming-Hsun Yang, Jwo-Yuh Wu, Tsang-Yi~Wang, Robert G. Maunder, and Rung-Hung Gau
\thanks{This work is sponsored by the Ministry of Science and Technology of Taiwan (MOST) under grants MOST 106 -2911-I-110-505 and MOST 107-2634-F-009-002. The work of J.-Y.~Wu is also supported by MOST Joint Research Center for AI Technology and AII Vista Healthcare.}
\thanks{M.-H. Yang, J.-Y.~Wu and R.-H.~Gau are with the Department of Electrical and Computer Engineering, National Chiao Tung University, Taiwan (e-mail: archenemy.cm00g@nctu.edu.tw; jywu@cc.nctu.edu.tw; runghunggau@g2.nctu.edu.tw).}
\thanks{T.-Y.~Wang is with the Institute of Communications Engineering, National Sun Yat-sen University, Kaohsiung, Taiwan 804 (e-mail: tcwang@mail.nsysu.edu.tw).}
\thanks{R. G. Maunder is with the Department of Electrical and Computer Engineering, University of Southampton, UK (e-mail: rm@ecs.soton.ac.uk).}}

\maketitle

\begin{abstract}
This paper proposes a fusion-based cooperative support identification scheme for distributed compressive sparse signal recovery via resource-constrained wireless sensor networks. The proposed support identification protocol involves: (i) local sparse sensing for economizing data gathering and storage, (ii) local binary decision making for partial support knowledge inference, (iii) binary information exchange among active nodes, and (iv) binary data aggregation for support estimation. Then, with the aid of the estimated signal support, a refined local decision is made at each node. Only the measurements of those informative nodes will be sent to the fusion center, which employs a weighted $\ell_1$-minimization for global signal reconstruction. The design of a Bayesian local decision rule is discussed, and the average communication cost is analyzed. Computer simulations are used to illustrate the effectiveness of the proposed scheme.
\end{abstract}


\section{Introduction}
Compressive sensing (CS) has provided a new signal processing paradigm whereby perfect/stable sparse signal recovery is provably true when using measurements sampled at rates below the Nyquist frequency \cite{BaraniukCS07,Eldar11}. Such a sub-Nyquist nature potentially economizes data gathering and storage; the reduction in measurement size can further facilitate efficient signal processing and conserve subsequent data transmission overheads. All these benefits have made CS pretty suitable for the design of resource-constrained wireless sensor networks (WSNs) \cite{Han13,JYWCL15,WimalajeewaICASSP13,Wimalajeewa17,LiWCL18}. Support identification is an important step in CS-based signal reconstruction, from both theoretical and application aspects \cite{Eldar11}. In the literature of CS-based WSNs, acquisition of a signal support estimate, or partial support knowledge, is crucial for the design of efficient distributed signal processing algorithms. For example, in the context of distributed sparse signal detection \cite{WimalajeewaICASSP13,Wimalajeewa17}, each local node first identifies a support and then projects its measurement onto the estimated signal subspace for noise reduction and reliable signal detection. For cost-aware WSNs, knowledge of a support estimate at the fusion center (FC) is needed to design sensor scheduling protocol towards energy reduction \cite{WChen:SP2016}. Regarding support identification in \cite{WimalajeewaICASSP13} and \cite{Wimalajeewa17}, each sensor node needs to gather a vector measurement of a sufficiently large size, and to conduct a CS-based reconstruction algorithm, such as orthogonal matching pursuit (OMP), for support estimation. This would place large data storage and computational burdens at the sensing devices.

To reduce the cost of support knowledge acquisition, in this paper we propose a fusion-based cooperative support identification and sparse signal reconstruction scheme. In the proposed approach, the $i$th sensor (i) employs a sparse sensing vector\footnote{Notably, sparse sensing vectors/matrices have been considered in the study of CS-based data acquisition and inference\cite{Gilbert10}.} $\Phib_{i}$ with support  $\mathcal{A}_{i}$ for data gathering, (ii) observes a scalar measurement (rather than a vector measurement) for partial support inference, and (iii) adopts 1-bit information exchange during the collaborative support identification phase. Notably, (i) and (ii) can economize data measurement and storage costs, whereas (iii) can reduce communication overhead. On the basis of (i) and (ii), we devise a binary local decision rule at each node to infer if the sensing vector support $\mathcal{A}_{i}$ overlaps with the desired signal support $\mathcal{T}$. If $\mathcal{A}_i\cap\mathcal{T}$ is judged to be nonempty, the sensor broadcasts a 1-bit message to all the other nodes (i.e., step (iii)), while otherwise keeping silent to conserve energy. Using the 1-bit messages received from all active nodes, each sensor forms a common support estimate $\hat{\mathcal{T}}$ by means of a simple counting rule. To the best of our knowledge, our study is the first in the literature which shows support identification can be realized by means of a cooperative binary decision-fusion based protocol. Once $\hat{\mathcal{T}}$ is available, only those nodes with $\mathcal{A}_i\cap\hat{\mathcal{T}}$ nonempty will forward their measurements to the FC for global signal reconstruction. The mean communication cost of the proposed scheme, which involves 1-bit information exchange for cooperative support identification and real-valued data transmission for global signal reconstruction, is analyzed. To fully exploit knowledge about $\hat{\mathcal{T}}$, the FC employs the weighted $\ell_1$-minimization algorithm \cite{CandèsBoyd08} for global signal reconstruction, with the weighting coefficients determined by $\hat{\mathcal{T}}$. Simulation results show that the proposed scheme outperforms the conventional method, which activates all the sensor nodes with real-valued data transmission, at a lower communication cost. 

\section{System model}

We consider a WSN, in which $M$ sensor nodes are coordinated by a FC to collaboratively estimate a $K$-sparse signal $\mathbf{s}\in\mathbb{R}^N$ with unknown support $\mathcal{T}\subset\{1,\ldots,N\}$ ($|\mathcal{T}|=K\ll N$). The $i$th sensor node makes a scalar observation obeying the following model
\begin{align}\label{eq:signal_model}
 y_{i}=\Phib_{i}^{T}\mathbf{s}+v_{i}, \quad 1\leq i\leq M,
\end{align}
where $y_{i}\in\mathbb{R}$ is the scalar measurement, $\Phib_{i}\in\mathbb{R}^N$ is a $K_c$-sparse sensing vector with support $\mathcal{A}_{i}\subset\{1,\ldots,N\}$ ($|\mathcal{A}_{i}|=K_c$), and $v_i\in\mathbb{R}$ is the observation noise assumed to be independent and identically distributed (i.i.d.) zero-mean Gaussian with variance $\sigma_v^2$, i.e., $v_i\sim\mathcal{N}(0,\sigma_v^2)$. The considered model can find applications in, e.g., cooperative wideband spectrum sensing in cognitive radio \cite{Meng11,Liu:Globecom2011}, in which networked cognitive users and an FC collaboratively estimate/detect a common primary user's signal occupying only a few (but unknown) frequency bands.

Thanks to the sparse nature of the unknown signal $\mathbf{s}$ and sensing vectors $\Phib_{i}$, \eqref{eq:signal_model} can be rewritten as
\begin{align}\label{eq:signal_model2}
    y_{i}=\left\{\begin{array}{ll}             \Phib_{i}^{T}\mathbf{s}+v_i, & \mbox{ if  $\mathcal{T}\cap\mathcal{A}_i\neq\emptyset$};\\
             v_i, & \mbox{ if  $\mathcal{T}\cap\mathcal{A}_i=\emptyset$, }\end{array}
             \right.
\end{align}
which in turn enables us to infer some partial knowledge about the signal support $\mathcal{T}$ at $i$th node. For instance, when the power of the noise $v_i$ is very small, certain elements in $\mathcal{A}_i$ shall be included in $\mathcal{T}$ if $|y_i|$ is not close to $0$, whereas all elements in $\mathcal{A}_i$ can be precluded from $\mathcal{T}$ whenever $|y_i|\approx 0$. Therefore, by exploiting such prior information conveyed by $y_i$ about the unknown signal support, this paper proposes a fusion-based cooperative support and signal reconstructing scheme. The following assumptions are made in the sequel.
\begin{assumption}\label{as:random_signal_support}
The signal support $\mathcal{T}$ is uniformly drawn from the collection $\mathbf{\Omega}_K:=\{\mathcal{T}_1,\cdots,\mathcal{T}_{C_K^N}\}$ of all $C_K^N$ possible sparsity pattern sets, where $\mathcal{T}_j\subset\{1,\ldots,N\}$ with $|\mathcal{T}_j|=K$ and $\Pr[\mathcal{T}_j]=1/C_K^N$.
\end{assumption}
\begin{assumption}\label{as:signal_distribution}
The nonzero entries of $\mathbf{s}$, say $s_k$, for $k\in\mathcal{T}$, are i.i.d. with $s_k\sim\mathcal{N}(0,\sigma_s^2)$, and are independent of the observation noise $v_i$'s.
\end{assumption}
\begin{assumption}\label{as:kc-sparse-sensing-vector}
For each $1\leq i\leq M$, the sensing vector $\mathbf{\Phi}_i$ is binary with $K_c$ nonzero entries, i.e., $\phi_{ij}\in\{+1,-1\}$ for $j\in\mathcal{A}_i$, and $|\mathcal{A}_i|=K_c$.
\end{assumption}
\begin{assumption}\label{as:random_sensing_support}
For each $1\leq i\leq M$, the sensing vector support $\mathcal{A}_i$ is uniformly drawn from the collection $\mathbf{\Omega}_{K_c}=\{\mathcal{A}_{i,1},\cdots,\mathcal{A}_{i,C_{K_c}^N}\}$ of all $C_{K_c}^N$ possible sparsity pattern sets, where $\mathcal{A}_{i,j}\subset\{1,\ldots,N\}$ with $|\mathcal{A}_{i,j}|=K_c$ and $\Pr[\mathcal{A}_{i,j}]=1/C_{K_c}^N$.
\end{assumption}
\begin{assumption}\label{as:FC-knows-sensing-vector}
The sensing vectors $\mathbf{\Phi}_i$'s, $1\leq i\leq M$, are known at the FC, whereas only their supports $\mathcal{A}_i$'s, $1\leq i\leq M$, are known at each sensor.
\end{assumption}

{\em Remark:} The uniform support location made in Assumption 1 is widely used in the literature of CS signal detection and estimation. This assumption is typically true in the cooperative spectrum sensing scenario, in which no prior knowledge about the frequency bands occupied by the primary user is available to the cognitive users. Assumption 2 regarding Gaussian signal entries is also quite standard in study of CS (e.g., \cite{Wimalajeewa2012,Mishra2016,Kailkhura17,Baron10}); related applications can also be found in spectrum sensing when the primary user adopts OFDM modulation\footnote{OFDM symbol is the inverse Fourier transform of independent finite-alphabet sources symbols, and is approximately Gaussian distributed especially when the number of sub-carriers is large \cite{Mishra2016}.}. Binary sparse sensing considered in Assumption 3 can be seen in, e.g., bio-medical imaging \cite{Mamaghanian11}, for reducing the data storage cost and execution time. Meanwhile, on account of the uniform assumption on the signal support distribution, a natural and reasonable rule for generating the sparse sensing vector supports is likewise the uniform distribution (Assumption 4). Finally, Assumption 5 is valid in scenarios such as cooperative spectrum sensing and source localization, in which network-wide knowledge of sensing vectors (either the full sparse sensing vectors or just their supports) can be acquired during the system built-up phase.

\section{Cooperative Support Identification}

\subsection{Proposed Protocol}
\textbf{Step I: Local Partial Support Inference}
\begin{itemize}
    \item Using the scalar observation $y_i$, the $i$th sensor node first makes its local decision $u_i$ to infer whether the desired sparse signal $\mathbf{s}$ lies in its sensing region or not, i.e.,
    \begin{align}\label{eq:local_support_estimation}
    u_{i}\left(y_i\right)=\left\{\begin{array}{ll}
             1, & \mbox{ if }
            \mathcal{T}\cap\mathcal{A}_i\neq\emptyset \mbox{ is decided};\\
             0, & \mbox{ otherwise. }\end{array}
             \right.
    \end{align}
    \item Afterwards, sensors with $u_i=1$ broadcast their binary local decisions to all the other nodes, while those with $u_i=0$ keep silent to conserve energy.
\end{itemize}

\textbf{Step II. Fusion for Support Identification}
\begin{itemize}
    \item Upon receiving the binary decisions $u_i$'s from the active nodes, each sensor computes for each index $n\in\{1,\dots,N\}$ the ``relative frequency'' that $n$ is activated during the sparse sensing process, namely,
	\begin{align}\label{eq:relative_freq}		    w\left(n\right)=\frac{\sum_{i\in\mathcal{I}}1\left\{ n\in\mathcal{A}_i\right\}}{\sum_{1\le i\le M}1\left\{ n\in\mathcal{A}_i\right\}},
		\end{align}
    where $1\{\cdot\}$ is an indicator function, and $\mathcal{I}\triangleq\{i|u_{i}(y_i)=1\}\subset\{1,\ldots,M\}$ is the active node index set during the support identification phase.
    \item Sort the values of $w(n)$ as $ w(n_1)\ge w(n_2)\ge\cdots\ge w(n_N)$. The proposed support estimate is obtained as
    \begin{align}\label{eq:That}
		   \hat{\mathcal{T}}(Z)=\left\{ n_1, n_2,\dots, n_Z\right\},
		\end{align}
	where $Z$ is an integer with $K\le Z\le N$. Note that $\hat{\mathcal{T}}(Z)$ is known to each sensor node.
	\item Nodes with $\mathcal{A}_{i}\cap\hat{\mathcal{T}}(Z)\neq\emptyset$ forward their real-valued measurements to the FC, which employs a weighted $\ell_1$-minimization for global signal reconstruction as to be discussed later.
\end{itemize}

Some comments are in order.
\begin{enumerate}
    \item In the literature of CS for WSNs, support recovery is typically done via greedy based search, e.g., the OMP or subspace pursuit \cite{Han13}; this involves computing a series of orthogonal projections, or least squares solutions (matrix inversion). Implementation of OMP-based iterations on the sensor nodes (e.g., \cite{WimalajeewaICASSP13,Wimalajeewa17}) would thus require large computation and data storage costs. Leveraging sparse sensing and collaboration among sensor nodes, the proposed scheme offers a fundamentally different methodology for support identification free from the need of matrix computation. Indeed, our scheme relies solely on local binary decision making and exchange, followed by a simple binary decision fusion \eqref{eq:relative_freq}. This makes the proposed approach rather suitable for WSNs subject to limited data storage, computation, and communication resources.
    \item Knowledge of $\hat{\mathcal{T}}(Z)$ will be exploited at the FC for conducting weighted $\ell_1$-minimization based global signal recovery (see Section IV). As a result, the signal reconstruction performance depends crucially on the support estimation quality. It is noted that the proposed cooperative support identification scheme via local 1-bit decision making and cooperative decision fusion are subject to two types of error: (i) Misidentification: $n_j\notin\hat{\mathcal{T}}(Z)$ is decided but $n_j\in\mathcal{T}$ is true. (ii) False Alarm: $n_j\in\hat{\mathcal{T}}(Z)$ is decided but $n_j\notin\mathcal{T}$ is true. Among the two error types, false alarm causes support over-estimation and is less harmful. This is because the computed signal amplitude on the over-estimated support element will typically assume a small value, leading to just a slight increase in the global signal reconstruction error. On the contrary, misidentification will be more dominant because missed support elements (i.e., support underestimate) cause severe model mismatch, which will incur a large reconstruction error.
    \item The cardinality $Z$ of the proposed support estimate $\hat{\mathcal{T}}(Z)$ in \eqref{eq:That} is allowed to range from $K$ (the true support size) to $N$ (the ambient dimension). Different values of $Z$ will result in different degree of robustness against the two error types and, thus, different signal reconstruction performance. If $Z=K$, a false alarm is necessarily accompanied by a misidentification, resulting in model mismatch. Such a drawback can be resolved by setting $Z>K$. For example, if one chooses $Z=K+2$, the proposed scheme can accommodate up to two false alarms. Hence, increasing the value of $Z$ is expected to improve quality of signal recovery. However, in the extreme case $Z=N$, there is no prior support knowledge; accordingly, all sensors directly forward their real-valued measurements to the FC, and the weighted $\ell_1$-minimization based signal reconstruction is reduced to the conventional $\ell_1$-minimization scheme without weighting. In light of the above discussions, the best global signal reconstruction performance will be achieved when $K<Z<N$; this will be confirmed by our simulation study.
    \item Define $\mathcal{S}\triangleq\{i|\mathcal{A}_{i}\cap\hat{\mathcal{T}}\neq\emptyset\}$ to be the index set of the participating nodes during the signal reconstruction phase. Notably, $\mathcal{S}$ does not necessarily coincide with $\mathcal{I}$.
    \item Implementation of the proposed scheme requires the knowledge of the true support size $K$ (or an upper bound). In CS-based WSNs, support size estimation is commonly done by using the residual-based algorithms or cross-validation \cite{Choi17}, which is typically implemented at the FC during the training phase \cite{Choi17}. For the proposed distributed protocol, a simple thresholding based approach is as follows. A sensor node broadcasts a one-bit decision $d_i(y_i)=1$ if $|y_i|$ is above a certain threshold. A coarse support size estimate can be obtained at each node as $\hat{K}=|\cup_{i\in\mathcal{J}}\mathcal{A}_{i}|$, where $\mathcal{J}\subset \{1,\dots,M\}$ is the active node index set during such a “support-size estimation phase”. Detailed design of the threshold for accurate support size estimation is beyond the scope of this paper.  
\end{enumerate}

\subsection{Bayesian Local Decision Rule}
The proposed support identification rule \eqref{eq:That} relies on fusion of the local binary decisions $\{u_{i}(y_i)\}_{i\in\mathcal{I}}$ according to \eqref{eq:relative_freq}. Hence, the quality of support estimate depends crucially on the accuracy of $u_i(y_i)$'s. Motivated by this fact, we obtain $u_i(\cdot)$ by solving the following problem:
\begin{align*}
    (P1) &\quad\min_{u_i} \Pr\left( u_i=1,\mathcal{T}\cap\mathcal{A}_i=\emptyset\right)\\&\qquad\qquad+\Pr\left( u_i=0,\mathcal{T}\cap\mathcal{A}_i\neq\emptyset\right).
\end{align*}
With some manipulations, the optimal solution to Problem $(P1)$ in the form of \eqref{eq:local_support_estimation} is expressed as
\begin{align}\label{eq:local_support_estimation2}
    u^*_i\left(y_i\right)=\left\{\begin{array}{ll}
             1, & \mbox{ if $\frac{p\left(\left.y_i\right|\mathcal{T}\cap\mathcal{A}_i\neq\emptyset\right)}{p\left(\left.y_i\right|\mathcal{T}\cap\mathcal{A}_i=\emptyset\right)}>\frac{\pi_0}{\pi_1}$};\\
             0, & \mbox{ if $\frac{p\left(\left.y_i\right|\mathcal{T}\cap\mathcal{A}_i\neq\emptyset\right)}{p\left(\left.y_i\right|\mathcal{T}\cap\mathcal{A}_i=\emptyset\right)}\le\frac{\pi_0}{\pi_1}$},\end{array}
             \right.
\end{align}
where $p\left(\left.y_i\right|\mathcal{T}\cap\mathcal{A}_i\neq\emptyset\right)$ and $p\left(\left.y_i\right|\mathcal{T}\cap\mathcal{A}_i=\emptyset\right)$ are the conditional probability density functions of $y_i$, $\pi_0=\Pr(\mathcal{T}\cap\mathcal{A}_i=\emptyset)$ and $\pi_1=\Pr(\mathcal{T}\cap\mathcal{A}_i\neq\emptyset)$ are the \textit{a priori} probabilities. By Assumptions \ref{as:random_signal_support}, \ref{as:signal_distribution}, and \ref{as:kc-sparse-sensing-vector}, the likelihood ratio of the measurement $y_i$ can be derived as
\begin{align}\label{eq:LR_local}
    &L\left(y_i\right)\triangleq\frac{p\left(\left.y_i\right|\mathcal{T}\cap\mathcal{A}_i\neq\emptyset\right)}{p\left(\left.y_i\right|\mathcal{T}\cap\mathcal{A}_i=\emptyset\right)}=\notag\\
    &\sum_{j=1}^{\min(K,K_c)}P_j\sqrt{\frac{\sigma_v^2}{j\sigma_s^2+\sigma_v^2}}\exp{\left(\frac{j\sigma_s^2 y_i^2}{2\sigma_{v}^{2}\left(j\sigma_s^2+\sigma_v^2\right)}\right)},
\end{align}
where 
\begin{align}\label{eq:conditional_probability1_coe}
   P_j=\frac{C_j^{K_c}C_{K-j}^{N-K_c}}{\sum_{j'=1}^{\min(K,K_c)}C_{j'}^{K_c}C_{K-j'}^{N-K_c}}.
\end{align}
Clearly, the restriction of $L(\cdot)$ on $\mathbb{R}^{+}\cup\{0\}$, say $L_0=L|_{\mathbb{R}^{+}\cup\{0\}}$, is a bijection and thus the corresponding inverse function $L_0^{-1}(\cdot)$ exists. Using this property together with some manipulations, the optimal decision rule in \eqref{eq:local_support_estimation2} can be expressed as 
\begin{align}\label{eq:local_support_estimation3}
    u^*_i\left(y_i\right)=\left\{\begin{array}{ll}
             1, & \mbox{ if $|y_i|>\eta\triangleq L_0^{-1}\left(\frac{\pi_0}{\pi_1}\right)$};\\
             0, & \mbox{ otherwise},\end{array}
             \right.
\end{align}
where $\pi_0=\frac{C^{N-K_c}_{K}}{C^{N}_{K}}$ and $\pi_1=\frac{\sum_{j=1}^{\min(K,K_c)}C_{j}^{K_c}C_{K-j}^{N-K_c}}{C^{N}_{K}}$.

\subsection{Communication Cost Analysis}
Based on the estimated signal support $\hat{\mathcal{T}}$, the $i$th node forwards its real-valued measurement $y_i$ to the FC if its sensing vector support overlaps with the estimated signal support, and keeps silent when otherwise. Accordingly, the expected communication cost of the $i$th node can be written as
\begin{align}\label{eq:local_transmission_cost}
    \beta_i\triangleq \alpha_1\Pr\left(i\in\mathcal{I}\right)+ \alpha_2\Pr\left(i\in\mathcal{S}\right),
\end{align}
where $\alpha_1>0$ is the communication cost when the $i$th node is active during the support identification phase, i.e., transmitting $u_i=1$, and $\alpha_2>0$ is the cost when the $i$th node participates in global signal reconstruction and transmits its real-valued measurement $y_i$ to the FC. It is noted that, in general, transmitting a real-valued data requires a higher communication cost than a binary bit, and hence $\alpha_2>\alpha_1$ is assumed. We have the following theorem.
\begin{theorem}\label{th:local_sent_cost}
For a WSN with $M$ sensor nodes and $|\hat{\mathcal{T}}|=Z$, the average communication cost required by the proposed scheme is bounded above by
\begin{align}\label{eq:total_transmission_cost}
    \beta_{T} &\le M\alpha_1\left[2\pi_1\sum_{j=1}^{\min(K,K_c)}P_jQ\left(\frac{\eta}{\sqrt{j\sigma_s^2+\sigma_v^2}}\right)+\right.\notag\\
    &\left.2\pi_0Q\left(\frac{\eta}{\sigma_v}\right)\right]+M\alpha_2\left(1-\frac{C^{N-Z}_{K_c}}{C^N_{K_c}}\right),
\end{align}
where $Q$ is the standard $Q$-function, $P_j$ is defined in \eqref{eq:conditional_probability1_coe}, and $\eta$, $\pi_0$ and $\pi_1$ are defined in \eqref{eq:local_support_estimation3}.
\end{theorem}
\begin{IEEEproof}
Since $p(y_i)=\pi_0p(y_i|\mathcal{T}\cap\mathcal{A}_i=\emptyset)+\pi_1p(y_i|\mathcal{T}\cap\mathcal{A}_i\neq\emptyset)$, it can be verified that
\begin{align}\label{eq:proof2}
&\Pr\left(i\in\mathcal{I}\right)=\Pr\left(\left| y_i\right|>\eta\right)=\notag\\
&2\pi_0Q\left(\frac{\eta}{\sigma_v}\right)+2\pi_1\sum_{j=1}^{\min(K,K_c)}P_jQ\left(\frac{\eta}{\sqrt{j\sigma_s^2+\sigma_v^2}}\right).
\end{align}
The probability $\Pr(i\in\mathcal{S})$ in \eqref{eq:local_transmission_cost} can be expressed as
\begin{align}\label{eq:proof3}
&\Pr\left(i\in\mathcal{S}\right)=1-\Pr\left(\mathcal{A}_{i}\cap\hat{\mathcal{T}}=\emptyset\right)\notag\\
&\stackrel{(a)}{\le}1-\sum_{\mathcal{B}\in\mathbf{\Omega}_{\hat{\mathcal{T}}}}\frac{C^{N-Z}_{K_c}}{C^N_{K_c}}\Pr\left(\hat{\mathcal{T}}=\mathcal{B}\right)=1-\frac{C^{N-Z}_{K_c}}{C^N_{K_c}},
\end{align}
where $(a)$ follows from Assumption \ref{as:random_sensing_support} and the dependence
among elements in $\hat{\mathcal{T}}$ (observed from \eqref{eq:relative_freq}), and $\Omega_{\hat{\mathcal{T}}}$ is the sample space of the estimated support $\hat{\mathcal{T}}$. 
With \eqref{eq:local_transmission_cost}, \eqref{eq:proof2} and \eqref{eq:proof3}, it can be shown that the cost $\beta_i$ is constant for all $i$, and hence, \eqref{eq:total_transmission_cost} follows immediately. 
\end{IEEEproof}

\section{Global Sparse Signal Reconstruction}
\label{sect:reconstruction}
\subsection{Signal Model}
Let $\Phib=\begin{bmatrix}\Phib_{1} & \Phib_{2} & \cdots & \Phib_{M}\end{bmatrix}^{T}\in\mathbb{R}^{M\times N}$ be the sensing matrix. Collecting all measurements $\{y_i\}_{i\in\mathcal{S}}$ into a vector, the received signal model at FC is given by
\begin{align}
    \mathbf{y}_{\mathcal{S}}=\Phib_{\mathcal{S}}\mathbf{s}+\mathbf{v}_{\mathcal{S}},
\end{align}
where $\mathbf{y}_{\mathcal{S}}\in\mathbb{R}^{|\mathcal{S}|}$ consists of $\{y_i\}_{i\in\mathcal{S}}$, $\Phib_{\mathcal{S}}\in\mathbb{R}^{|\mathcal{S}|\times N}$ is obtained by retaining the rows of $\Phib$ indexed by $\mathcal{S}$, and $\mathbf{v}_{\mathcal{S}}\in\mathbb{R}^{|\mathcal{S}|}$ is the noise vector. With the aid of $\hat{\mathcal{T}}$, the estimated signal is obtained by solving the following
weighted $\ell_1$-minimization problem
\begin{align*}
    (P2) \qquad \min_{\mathbf{s}} \|\mathbf{Ws}\|_{1}, \quad \mbox{ s.t. } \|\mathbf{y}_{\mathcal{S}}-\Phib_{\mathcal{S}}\mathbf{s}\|_{2}<\epsilon
\end{align*}
where $\mathbf{W}=diag\{\omega_{1},\dots,\omega_{N}\}$ with $\omega_{k}\in[0,1]$ being the weighting coefficient assigned to the $k$th index, and $\epsilon>0$ specifies the error level. Following \cite{Friedlander12}, in this paper we assign $\omega_k$ a smaller value when $k\in\hat{\mathcal{T}}$, and a greater value, otherwise.

\begin{figure}[hbt]
\begin{center}
\includegraphics[scale=0.39]{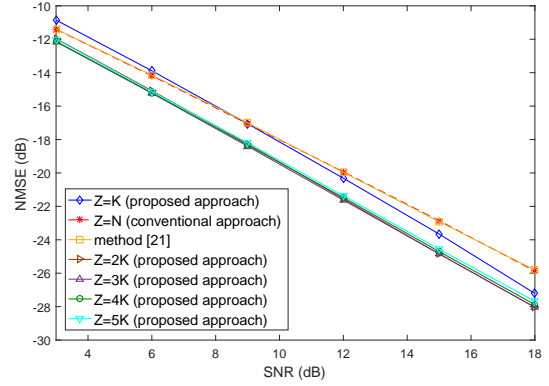}
\caption{NMSE as a function of SNR when $Z=K$, $2K$, $3K$, $4K$, $5K$, and $N$. }\label{fig:snr_vs_err} 
\end{center}
\end{figure}

\subsection{Coherence of Sparse Sensing Matrix $\Phib$}
On account of Assumptions 3-4, the following theorem shows that, with a very high probability, the scaled sparse sensing matrix $\sqrt{N/(K_{c}M)}\Phib$ satisfies the restricted isometry property (RIP) of order $K(\geq 2)$ with a small restricted isometry constant (RIC) $0<\delta_K<1$.
\begin{theorem}\label{th:rip_of_phi}
For every sparsity level $1\leq K\leq N$ and every $\delta\in(0,1)$, if
\begin{align}
    M\geq C\delta^{-2}K\log\left(\frac{eN}{K}\right),
\end{align}
the scaled sensing matrix $\sqrt{\frac{N}{K_cM}}\Phib$ satisfies the RIP of order $K$ with RIC $\delta_K\leq\delta$ with probability exceeding $1-2\exp(-c\delta^2M)$, where $C$ and $c$ are positive absolute constants.
\end{theorem}
\begin{IEEEproof}
See Appendix \ref{app.A}.
\end{IEEEproof}
Also, under the above RIP assumption and with \cite[Lemma 2.1]{CC14}, we have the following theorem.
\begin{theorem}\label{th:coherence_of_phi}
If the scaled sensing matrix $\sqrt{\frac{N}{K_cM}}\Phib$ satisfies the RIP of order $K(\geq2)$ with RIC $\delta_K$, the coherence $\mu_c$ of the sensing matrix $\Phib$ satisfies \begin{align}\mu_c=\max_{1\leq i\neq j\leq N}\frac{|<\mathbf{c}_i,\mathbf{c}_j>|}{\|\mathbf{c}_i\|_{2}\|\mathbf{c}_j\|_{2}}\leq\frac{\delta_K}{1-\delta_K},\end{align}
where $\mathbf{c}_i$ is the $i$th column of $\Phib$, $1\leq i\leq N$.
\end{theorem}
\begin{IEEEproof}
See Appendix \ref{app.B}.
\end{IEEEproof}

Hence, the coherence $\mu_c$ can be kept small with a high probability, thereby guaranteeing the robustness of the proposed collaborative sparse signal estimation scheme.

\begin{figure}[hbt]
\begin{center}
\includegraphics[scale=0.39]{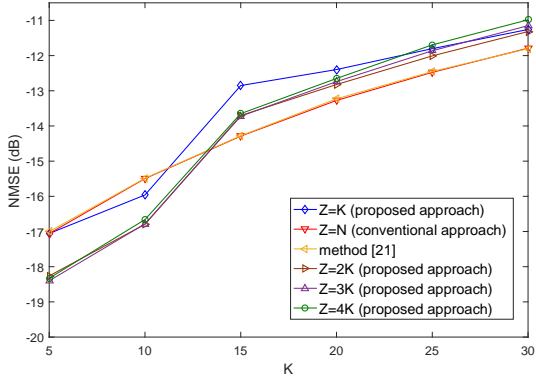}
\caption{NMSE as a function of $K$ when $Z=K$, $2K$, $3K$, $4K$, and $N$ (SNR=9 dB).}\label{fig:NMSE_vs_K}
\end{center}
\end{figure}

\section{Performance evaluation}
In this section, computer simulations are provided to demonstrate the effectiveness of the proposed scheme.  The ambient signal dimension is set to be $N=500$ and the network size is $M=350$. The number of non-zeros in the compression vectors $\mathbf{\Phi}_i$'s is $K_c=50$. The weighting coefficient $\omega_k$ is set to be $\omega_k=0.5$ if $k\in\hat{\mathcal{T}}$, whereas $\omega_k=1$ if $k\notin\hat{\mathcal{T}}$. The SNR of the local sensor measurement is defined as SNR$\triangleq E\{\mathbf{(\Phi}_i^{T}\mathbf{s})^2\}/E\{v_i^2\}$. The quality of signal recovery is evaluated by using the normalized mean square error (NMSE), defined as $\mbox{NMSE}\triangleq E\left\{\|\mathbf{s}-\hat{\mathbf{s}}\|^2/\|\mathbf{s}\|^2\right\}$, where $\hat{\mathbf{s}}$ is the reconstructed sparse signal at the FC. In the discussions below, the method in \cite{Wimalajeewa15}, which also addressed distributed sparse signal estimation via sparse measurement matrices, is used as the comparative scheme. The simulation results are obtained from $20000$ independent trials.

\begin{figure}[t]
\begin{center}
\includegraphics[scale=0.39]{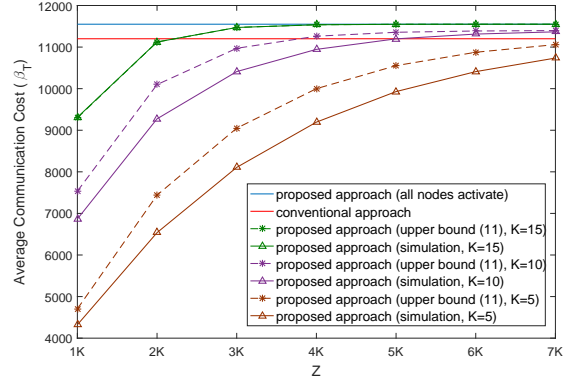}
\caption{Average communication cost $\beta_{T}$ as a function of $Z$ when $K=5$, $10$ and $15$ (SNR=9 dB). }\label{fig:m_vs_err}
\end{center}
\end{figure}

In the first example, we evaluate the proposed scheme with different value of size $Z$. For $K=5$, Fig. \ref{fig:snr_vs_err} plots the NMSE with respect to (w.r.t.) different SNR for $Z=K$, $2K$, $3K$, $4K$, $5K$, and $N$. As mentioned earlier, when $Z=N$, the proposed scheme reduces to the conventional CS approach, which activates all sensor nodes and utilizes standard $\ell_1$-minimization for signal reconstruction. The figure shows the NMSE performance of \cite{Wimalajeewa15} is very close to the conventional CS system, and the proposed scheme with $Z>K$ outperforms these two methods. Note that our method with $Z=2K$ achieves the lowest NMSE, confirming our discussions that the best value of $Z$ falls between the range from $K$ to $N$. For SNR=9 dB, Fig. \ref{fig:NMSE_vs_K} compares the NMSE for different sparsity level $K$. The figure shows the performances of all methods degrade as $K$ increases. The proposed scheme incurs larger NMSE as $K$ is above 14; this is because, as $K$ increases, support size over-estimation ($|\hat{\mathcal{T}}|=Z>K=|\mathcal{T}|$) becomes severe, resulting in undesirable error floor. To compare the required communication costs, we set $\alpha_1=1$ and $\alpha_2=32$. For SNR=9 dB, Fig. \ref{fig:m_vs_err} plots the average communication costs w.r.t. $Z$ for three sparsity levels $K=5, 10, 15$; both the theoretical upper bounds \eqref{eq:total_transmission_cost} and the simulated results are included. The blue curve depicts the baseline communication cost (equal to $M\alpha_1+M\alpha_2=11550$) of the proposed scheme that accounts for the communication during the support identification phase and the data transmission phase with all sensor activated. We observe the following: (i) the communication cost of our method increases with $K$ and $Z$; (ii) compared with the conventional CS method, the proposed scheme can reduce the cost when $K\le10$ and $Z\le4K$, but incurs more cost as $K$ and $Z$ increase since more sensors are activated. We note that the communication cost of the method in \cite{Wimalajeewa15} is large (equal to $MN\alpha_2=5600000$) because the protocol in \cite{Wimalajeewa15} involves a large amount of real-valued data transmission. 


\appendices
\section{Proof of Theorem \ref{th:rip_of_phi} }\label{app.A}
We will first prove that $\sqrt{\frac{N}{K_{c}}}\Phib_i$ is an isotropic sub-Gaussian random vector. Then, with the aid of Theorem 5.65 in \cite{Eldar11}, the assertion of Theorem \ref{th:rip_of_phi} immediately follows. The sub-Gaussianalty of $\sqrt{\frac{N}{K_{c}}}\Phib_i$ is established by the following lemma.
\begin{lemma}
Let $\mathbf{q}\in\mathbb{R}^N$ be a  $K_c$-sparse vector with support $\mathcal{T}_{\mathbf{q}}\subset\{1,\ldots,N\}$ uniformly drawn from the collection $\mathbf{\Omega}_{K_c}\triangleq\{\mathcal{T}_1,\cdots,\mathcal{T}_{C_{K_c}^N}\}$ of all $C_{K_c}^N$ possible sparsity patterns. The nonzero entries of $\mathbf{q}$ are assumed to be independent symmetric Bernoulli random variables, i.e., $q_i\in\{+1,-1\}$ with $\Pr(q_i=1)=\Pr(q_i=-1)=1/2$ for $i\in\mathcal{T}_{\mathbf{q}}$. Then $\frac{1}{\sqrt{\rho}}\mathbf{q}$ is an isotropic sub-Gaussian random vector with constant $\alpha=\Bar{c}/\sqrt{\rho}$, where $\Bar{c}>0$ is a constant and $\rho=K_c/N$.
\end{lemma}
\begin{IEEEproof}
First, for each $1\leq i\leq C_{K_c}^N$, straightforward manipulations show that the conditional expectation $E\{\mathbf{q}\mathbf{q}^T|\mathcal{T}_{\mathbf{q}}=\mathcal{T}_{i}\}=\mathbf{C}_i$, where $\mathbf{C}_i\in\mathbb{R}^{N\times N}$ is a diagonal matrix with $[\mathbf{C}_i]_{jj}=1$  if $j\in\mathcal{T}_{i}$  and $[\mathbf{C}_i]_{jj}=0$ when otherwise. Then, the \textit{second moment matrix} of $\frac{1}{\sqrt{\rho}}\mathbf{q}$ can be obtained as follows,
\begin{align}\label{eq:rip_proof1}
    &E\left\{\frac{1}{\sqrt{\rho}}\mathbf{q}\frac{1}{\sqrt{\rho}}\mathbf{q}^T\right\}=\frac{1}{\rho}E\left\{\mathbf{q}\mathbf{q}^T\right\}\notag\\
    &=\frac{1}{\rho}\sum_{i=1}^{C_{K_c}^N}E\left\{\left.\mathbf{q}\mathbf{q}^T\right|\mathcal{T}_{\mathbf{q}}=\mathcal{T}_{i}\right\}\Pr\left(\mathcal{T}_{\mathbf{q}}=\mathcal{T}_{i}\right)\notag\\
    &=\frac{1}{\rho C_{K_c}^N}\sum_{i=1}^{C_{K_c}^N}\mathbf{C}_i\notag\\
    &=\mathbf{I}_{N}.
\end{align}
Hence, by definition, the random vector $\frac{1}{\sqrt{\rho}}\mathbf{q}$ is isotropic. To prove $\frac{1}{\sqrt{\rho}}\mathbf{q}$ is a sub-Gaussian random vector, we need to check that, for every $\mathbf{a}\in\mathbb{R}^N$, the inner product $<\frac{1}{\sqrt{\rho}}\mathbf{q},\mathbf{a}>$ is sub-Gaussian random variable. To see this, let $t\geq0$ and then we have
\begin{align}\label{eq:rip_proof2}
    &\Pr\left(\left|<\frac{1}{\sqrt{\rho}}\mathbf{q},\mathbf{a}>\right|>t\right)\notag\\
    &=\sum_{i=1}^{C_{K_c}^N}\Pr\left(\left.\left|<\frac{1}{\sqrt{\rho}}\mathbf{q},\mathbf{a}>\right|>t\right|\mathcal{T}_{\mathbf{q}}=\mathcal{T}_{i}\right)\Pr\left(\mathcal{T}_{\mathbf{q}}=\mathcal{T}_{i}\right)\notag\\
    &=\sum_{i=1}^{C_{K_c}^N}\Pr\left(\left|\sum_{j\in\mathcal{T}_{i}}\frac{1}{\sqrt{\rho}}q_{j}a_{j}\right|>t\right)\Pr\left(\mathcal{T}_{\mathbf{q}}=\mathcal{T}_{i}\right)\notag\\
    &\stackrel{(a)}{\leq}\sum_{i=1}^{C_{K_c}^N}\left[e\exp\left(-\frac{\rho ct^2}{\|\mathbf{a}_{\mathcal{T}_{i}}\|_2^2}\right)\right]\Pr\left(\mathcal{T}_{\mathbf{q}}=\mathcal{T}_{i}\right)\notag\\
    &\leq\sum_{i=1}^{C_{K_c}^N}\left[e\exp\left(-\frac{\rho ct^2}{\|\mathbf{a}\|_2^2}\right)\right]\Pr\left(\mathcal{T}_{\mathbf{q}}=\mathcal{T}_{i}\right)\notag\\
    &=e\exp\left(-\frac{\rho ct^2}{\|\mathbf{a}\|_2^2}\right),
\end{align}
where (a) holds due to the fact that $q_j$'s are independent symmetric Bernoulli random variables for all $j\in\mathcal{T}_{i}$ and thus, by Proposition 5.10 in \cite[Chap. 5]{Eldar11}, the inequality $\Pr\left(\left|\sum_{j\in\mathcal{T}_{i}}\frac{1}{\sqrt{\rho}}q_{j}a_{j}\right|\geq t\right)\leq e\exp\left(-\frac{\rho ct^2}{\|\mathbf{a}_{\mathcal{T}_{i}}\|_2^2}\right)$ is valid, where $\mathbf{a}_{\mathcal{T}_{i}}\in\mathbb{R}^{K_c}$ is obtained by keeping the entries of $\mathbf{a}$ indexed by $\mathcal{T}_{i}$ and $c>0$ is an absolute constant. Inequality \eqref{eq:rip_proof2} shows that the random vector $\frac{1}{\sqrt{\rho}}\mathbf{q}$ is sub-Gaussian random vector and the corresponding sub-Gaussian norm is bounded above by $\Bar{c}/\sqrt{\rho}$, where $\Bar{c}>0$. Therefore, $\frac{1}{\sqrt{\rho}}\mathbf{q}$ is an isotropic sub-Gaussian random vector in $\mathbb{R}^N$ with constant $\Bar{c}/\sqrt{\rho}$.
\end{IEEEproof}

Based on Lemma 1, the assertion of Theorem \ref{th:rip_of_phi} immediately follows the next lemma.

\textit{Lemma 2 \cite[Theorem 5.65]{Eldar11}:} Let $\mathbf{A}$ be an $M\times N$ sub-Gaussian random matrix, which each row is independent isotropic sub-Gaussian random vector. Then for every sparsity level $1\leq K\leq N$ and every $\delta\in(0,1)$, if
\begin{align}
    M\geq C\delta^{-2}K\log\left(\frac{eN}{K}\right),
\end{align}
the scaled matrix $\sqrt{\frac{1}{M}}\mathbf{A}$ satisfies the RIP of order $K$ with RIC $\delta_K\leq\delta$ with probability exceeding $1-2\exp(-c\delta^2M)$, where $C$ and $c$ are positive absolute constants and depend only on the sub-Gaussian norm of the rows of $\mathbf{A}$.

\section{Proof of Theorem \ref{th:coherence_of_phi} }\label{app.B}
If $\sqrt{\frac{N}{K_{c}M}}\Phib$ has RIP, then for any $1\leq i\neq j\leq N$, we have
\begin{align}\label{eq:commfent5_proof1}
    \frac{\left|\left<\mathbf{c}_{i},\mathbf{c}_{j}\right>\right|}{\left\|\mathbf{c}_{i}\right\|_{2}\left\|\mathbf{c}_{j}\right\|_{2}}&=\frac{\left|\left<\sqrt{\frac{N}{K_{c}M}}\Phib\mathbf{e}_{i},\sqrt{\frac{N}{K_{c}M}}\Phib\mathbf{e}_{j}\right>\right|}{\left\|\sqrt{\frac{N}{K_{c}M}}\Phib\mathbf{e}_{i}\right\|_{2}\left\|\sqrt{\frac{N}{K_{c}M}}\Phib\mathbf{e}_{j}\right\|_{2}}\notag\\
    &\stackrel{(a)}{\leq}\frac{\delta_{2}\left\|\mathbf{e}_{i}\right\|_2\left\|\mathbf{e}_{j}\right\|_2}{\left\|\sqrt{\frac{N}{K_{c}M}}\Phib\mathbf{e}_{i}\right\|_{2}\left\|\sqrt{\frac{N}{K_{c}M}}\Phib\mathbf{e}_{j}\right\|_{2}}\notag\\
    &\stackrel{(b)}{\leq}\frac{\delta_{2}\left\|\mathbf{e}_{i}\right\|_2\left\|\mathbf{e}_{j}\right\|_2}{\left(1-\delta_{K}\right)\left\|\mathbf{e}_{i}\right\|_2\left\|\mathbf{e}_{j}\right\|_2}\notag\\
    &=\frac{\delta_{2}}{1-\delta_{K}}\notag\\
    &\stackrel{(c)}{\leq}\frac{\delta_{K}}{1-\delta_{K}},
\end{align}
where (a) follows from Lemma 2.1 in \cite{CC14} and (b) holds because $\sqrt{\frac{N}{K_{c}M}}\Phib$ satisfies RIP with constant $\delta_K$, and (c) is true since $\delta_2\leq\delta_K$ for $K\geq2$. Therefore, $\mu_c=\max_{1\leq i\neq j\leq N}\frac{\left|\left<\mathbf{c}_{i},\mathbf{c}_{j}\right>\right|}{\left\|\mathbf{c}_{i}\right\|_{2}\left\|\mathbf{c}_{j}\right\|_{2}}\leq\frac{\delta_{K}}{1-\delta_{K}}$.


\end{document}